\documentclass[sigconf, nonacm]{acmart}

\makeatletter
\def\@ACM@checkaffil{}
\makeatother

\usepackage{amsmath,amsfonts}
\usepackage{algorithmic}
\usepackage{graphicx}
\usepackage{textcomp}
\usepackage{xcolor}
\usepackage[table]{xcolor}
\definecolor{matmulonly}{RGB}{221,235,247}
\definecolor{mvmulonly}{RGB}{226,239,218}
\usepackage{todonotes}
\usepackage{booktabs}
\usepackage{colortbl}
\usepackage{subcaption} 
\usepackage{listings}
\usepackage{multirow}
\usepackage[table]{xcolor}
\usepackage{siunitx}

\definecolor{lightgray}{gray}{0.93}
\def\BibTeX{{\rm B\kern-.05em{\sc i\kern-.025em b}\kern-.08em
    T\kern-.1667em\lower.7ex\hbox{E}\kern-.125emX}}
\graphicspath{{figures/}}

\begin{document}

\title{GPU Goldmines: Tunable WebGPU Kernels for Edge LLM Inference}

\author{Rithik Sharma}
\affiliation{%
  \institution{UC Santa Cruz}
}

\author{Reese Levine}
\affiliation{%
  \institution{UC Santa Cruz}
}

\author{Zheyuan Chen}
\affiliation{%
  \institution{UC Santa Cruz}
}

\author{Ananthajit Srikanth}
\affiliation{%
  \institution{EPFL}
}

\author{Yanwen Xu}
\affiliation{%
  \institution{UC Santa Cruz}
}

\author{Tyler Sorensen}
\affiliation{%
  \institution{Microsoft Research}
  \institution{UC Santa Cruz}
}

\begin{abstract}

As demand for LLM inference grows and consumer GPUs grow more capable, edge deployment is becoming increasingly attractive. WebGPU is compelling in this setting because it offers cross-platform portability and browser accessibility. However, these same strengths also make high performance difficult to achieve, as an implementation that performs well in one context may not perform well in another. 

To address this, we present GPU Goldmines, a framework for specifying, executing, and tuning WebGPU kernels. We instantiate GPU Goldmines with kernels for key LLM building blocks: matrix multiplication, matrix-vector multiplication, and flash attention. We characterize performance across diverse devices, shapes, and browsers, and use the resulting database to accelerate subsequent autotuning and configuration selection. On shapes drawn from Llama and Qwen models, our matrix multiplication kernels outperform ONNX Runtime and llama.cpp WebGPU kernels by $7.0\times$ and $1.41\times$ geometric mean, respectively, and deliver end-to-end improvements of $1.48\times$ in the prefill stage of llama.cpp for chat workloads.
\end{abstract}

\maketitle

\section{Introduction}
\label{sec:intro}

% Motivate for edge LLM is practical
Large language model inference on consumer devices presents an exciting and practical new deployment opportunity~\cite{lee2019ondevice}. Running models locally reduces latency and removes dependencies on datacenters~\cite{zheng2025edge, husom2025edgeai, li2024mobilellm}. WebGPU is poised to be a key enabler of this by exposing GPU acceleration inside the browser without requiring extra installations or platform-specific software stacks~\cite{webgpu}. Systems such as WebLLM~\cite{ruan2024webllm}, ONNX Runtime Web~\cite{onnx}, and llama.cpp's WebGPU backend~\cite{llama_cpp, levine2026llamaweb} already provide initial (and relatively performant) LLM implementations for WebGPU.

%Motivating for performance is hard
That portability, however, requires spanning a remarkably diverse target space. WebGPU kernels may execute on integrated, mobile, and discrete GPUs from different vendors, each with architectural characteristics that impact performance, such as subgroup (warp) size, the bandwidth and latency of the memory hierarchy, and arithmetic throughput~\cite{cudadma}. The software stack is similarly diverse; depending on the platform and browser, WebGPU kernels are compiled and run using different implementations and backend APIs, e.g., DirectX, Vulkan, and Metal~\cite{directx-specs, vulkan1.3, apple_metal}. As a result, a kernel implementation that performs well in one context may perform poorly in another. In short, while WebGPU provides functional portability, performance portability remains a difficult problem, as previous research has shown~\cite{sorensen2019onesize, price2017perfport, zhang2013perfport}.

%Summarize related works
One way to address performance portability is to make kernels tunable. Rather than fixing a single kernel implementation, a tunable kernel exposes a semantics-preserving family of implementations parameterized by performance-relevant choices. Many choices are explicit scalar parameters, such as workgroup size and tile dimensions. Autotuning frameworks can then search this space empirically to identify high-performing instantiations for a target platform. Systems such as OpenTuner~\cite{ansel2014opentuner}, CLTune~\cite{nugteren2015cltuner}, and KernelTuner~\cite{kerneltuner} follow this model, using advanced heuristics to search these spaces efficiently. Tuning is also not limited to assigning values to a fixed set of parameters. In schedule-based systems such as Halide~\cite{ragankelley2013halide} and TVM~\cite{chen2018learning}, the search space consists of program transformations, such as tiling, fusion, and reordering. This raises the level of abstraction and expresses richer optimization spaces. Moving even higher up the abstraction hierarchy, AI kernel optimization has shown promise as an even more advanced form of tuning~\cite{ouyang2025kernelbenchllmswriteefficient, wen2025multikernelbench, li2025tritonbenchbenchmarkinglargelanguage}.

%Highlight shortcomings
However, these tuning approaches suffer from two problems: (1) they rely on immediate empirical feedback from the device being tuned on, with little-to-no ability to translate performance from one device to another, and (2) they lack centralized stores of tuning data that can be utilized to provide optimal parameters when running on known devices or reduce the tuning time needed on previously unseen devices.
%However, because these tuning approaches rely on immediate empirical feedback, they are designed for narrow targets, whereas WebGPU’s strength lies in its broad and diverse deployment space. 
This mismatch is critical for LLM inference, where execution is concentrated in a small number of performance-critical kernels. Because these kernels are invoked repeatedly, tuning them for one device or browser stack and running them in another can directly degrade important application performance, including time-to-first-token and tokens-per-second. Manual inspection of existing WebGPU LLM frameworks suggests that this is currently the case, as existing systems provide limited specialization across devices and browsers (Sec.~\ref{sec:e2e}).

\subsection{GPU Goldmines}
\label{sec:intro-gpugoldmines}

To address this challenge, we present \textit{GPU Goldmines}, a framework for tuning performance-critical WebGPU kernels for LLM inference. GPU Goldmines performs tuning in a specific execution context, but retains the resulting measurements in a centralized performance database. By systematically exploring these kernel tuning spaces across devices, browsers, and workload shapes, the database provides the basis for three capabilities: (1)~deep characterization across WebGPU’s broad target space, (2)~accelerating subsequent tuning runs by reusing prior measurements, and (3)~constructing decision procedures that map execution contexts to high-performing parameter settings. Because GPU Goldmines runs entirely in the browser, it requires no installation or platform-specific dependencies, allowing any compatible device to contribute benchmarks and allowing the database to grow incrementally as the WebGPU ecosystem evolves.

\paragraph{Specifying, Executing, and Recording}
GPU Goldmines provides an interface for defining tunable kernels, along with tools for systematically executing them and recording their performance. A tunable kernel is defined by a parameterized implementation, exposed tuning parameters, validity constraints, and input properties defining the execution context. From this specification, the system can generate and execute concrete kernel instances by either exhaustively exploring the valid parameter space or searching it heuristically. All measurements are written back to a centralized database, allowing tuning results to accumulate across runs and devices. To ensure that these measurements are reliable, GPU Goldmines includes built-in support for warm-up executions, repeated timing iterations, and correctness validation.

\paragraph{Tunable LLM Kernels and Characterization}
GPU Goldmines targets three performance-critical LLM routines: matrix multiplication, matrix-vector multiplication, and flash attention. For each, we build tunable WebGPU kernel families by drawing from existing WebGPU implementations, portable GPU libraries, and well-established implementation strategies, and by exposing performance-critical parameters such as hierarchical tile sizes and workgroup dimensions. We instantiate these families across multiple precisions and data formats, including FP32, FP16, and quantized variants, and evaluate them on shapes drawn from representative Llama and Qwen inference workloads spanning both short sequence length conversational settings and longer sequence length summarization input shapes.

\begin{figure}[t]
\centering
\includegraphics[width=\linewidth]{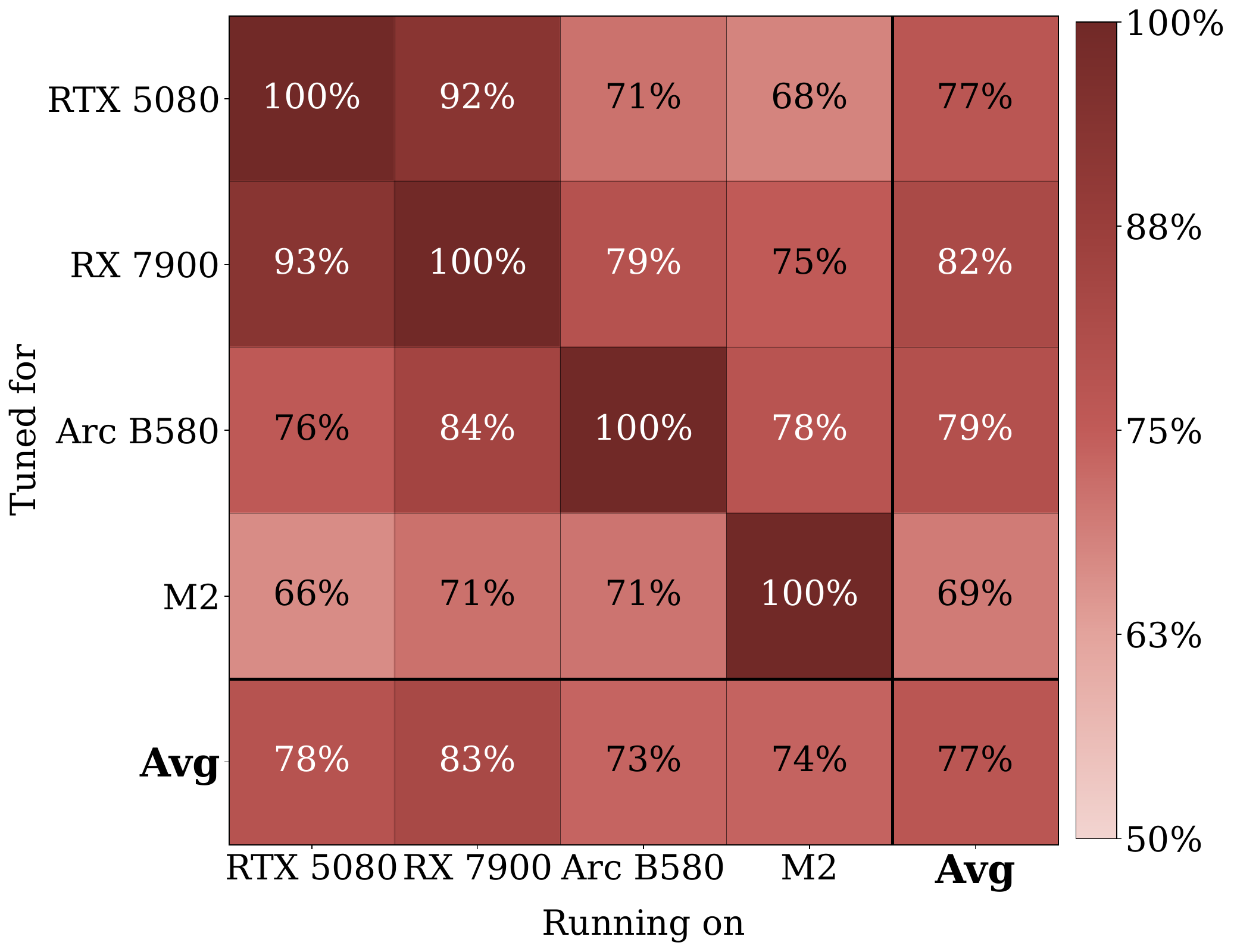}\caption{Performance portability of matmul-tiled across 4 devices and 12 LLM inference sizes. Each cell shows the percentage of device-optimal performance retained when transferring a configuration tuned on one device to another.}\label{fig:perfport}\end{figure}

For a representative subset of these kernel families, we exhaustively explore the tuning space across four diverse devices: an RTX 5080, an RX 7900XT, an Arc B580, and an M2 GPU (details given in Table~\ref{tab:gpus}). This enables deep characterization of how performance varies across implementations and execution contexts. The resulting data reveals substantial portability gaps. For example, Figure~\ref{fig:perfport} shows a performance portability heatmap for FP16 matrix multiplication, where a configuration tuned for one device achieves only 77\% of its potential performance on another device on average. At the same time, matrix-vector multiplication is markedly less sensitive to tuning parameters (Sec.~\ref{sec:perf}), motivating our focus on \textit{matmul} and prefill performance. Although these exhaustive studies are expensive, they also provide a strong seed for the centralized database, which can then be reused to accelerate subsequent tuning runs.

\paragraph{Parameter Selection and Accelerated Tuning}
GPU Goldmines uses its centralized database not only to characterize WebGPU performance, but also to make future tuning cheaper and more effective. For devices already represented in the database, we derive simple parameter selection policies that map execution contexts and input shapes to high-performing configurations, reaching an average of 90\% of hyper-tuned performance for fixed input sizes. For previously unseen devices, database-derived configurations provide strong warm starts for heuristic search, enabling high-performing parameters to be identified 52.4\% faster than tuning from scratch. For previously unseen input shapes, we evaluate transfer methods that reason over tensor dimensions: a distance-based selector and a Bayesian approach that uses prior observations to guide search. These results demonstrate that performance data can be reused not only to analyze WebGPU kernels, but also to predict, initialize, and accelerate tuning in new deployment contexts.

\paragraph{Evaluation}
Finally, we evaluate GPU Goldmines at both the kernel and application levels across multiple experimental settings spanning 8 GPUs. At the kernel level, our tuned configurations outperform the WebGPU kernels used by existing LLM systems, delivering geometric mean speedups of $1.41\times$ over llama.cpp kernels and $7.0\times$ over ONNX Runtime kernels. The large gap over ONNX Runtime reflects two limitations of its current WebGPU kernels: (1)~lack of hardware matrix unit support, and (2)~tuning for long sequence lengths, which transfer poorly to chat workloads.
At the application level, we integrate our configuration selection techniques into llama.cpp and evaluate end-to-end inference on standard LLM benchmarks using representative Llama and Qwen models. We observe an average end-to-end prefill speedup of $1.48\times$ across these workloads (Sec.~\ref{sec:llama-e2e}).

% they do not target hardware matrix acceleration units, even when such units are available, and they appear to be tuned for large sequence lengths, which transfer poorly to the shorter sequence lengths common in chat workloads.

\paragraph{Contributions}
In summary, our contributions are:
\begin{itemize}
\itemsep -0.5mm
  \item GPU Goldmines, a browser-based framework for specifying tunable WebGPU kernels, systematically executing valid configurations, and recording performance measurements (Sec.~\ref{sec:framework}).
  \item Tunable WebGPU kernel families for three performance-critical LLM routines: matrix multiplication, matrix-vector multiplication, and flash attention, together with a large-scale characterization of their performance across devices, browsers, precisions, and workload shapes (Sec.~\ref{sec:shaders}, Sec.~\ref{sec:perf}).
  \item Data-driven methods that reuse the database to characterize portability, derive configuration selection policies, and accelerate tuning in new contexts (Sec.~\ref{sec:database}).
  \item Kernel-level and end-to-end evaluation showing that our tuned configurations outperform existing WebGPU LLM kernels, including $7.0\times$ over ONNX Runtime and $1.41\times$ over llama.cpp, with average end-to-end prefill speedups of $1.48\times$ in llama.cpp (Sec.~\ref{sec:e2e}).
\end{itemize}

\noindent
Code and data will be released as open source after publication, along with a public deployment of GPU Goldmines that can continue to grow and evolve.

\begin{figure*}[t]
\centering
\includegraphics[width=0.90\textwidth]{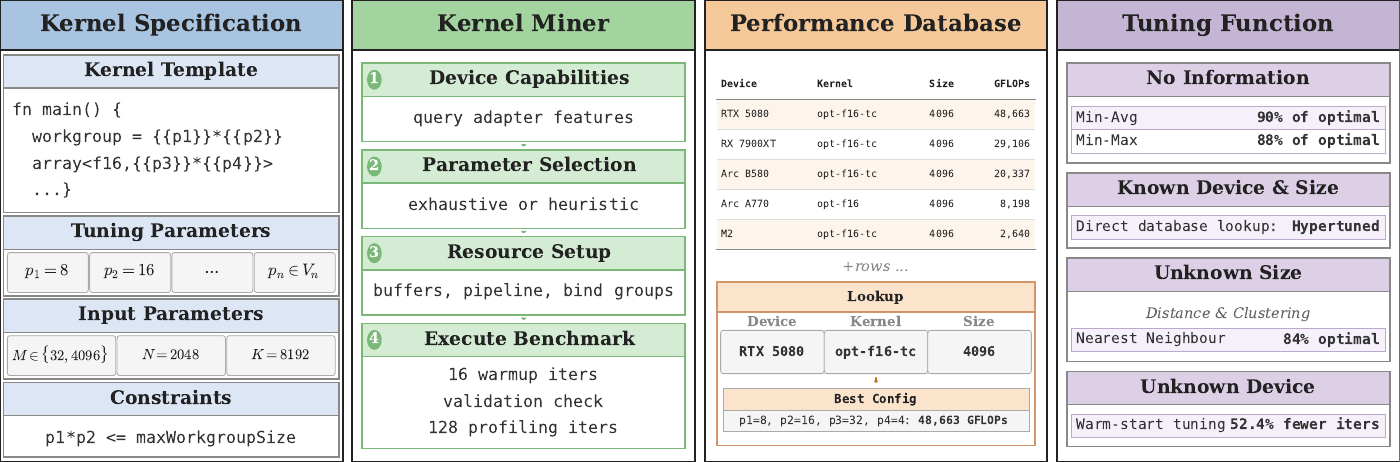}
\caption{Overview of the GPU Goldmines toolsuite. A tunable kernel specification is instantiated and benchmarked by the kernel miner on the target device, with results stored in a centralized performance database. The tuning function uses this database to select configurations for four deployment scenarios. Tuned configurations are evaluated against ONNX Runtime, SurfGrad, and integrated into llama.cpp for end-to-end inference.}
\label{fig:overview}
\end{figure*}

\section{Background}
\label{sec:background}

\subsection{LLM Inference}

Many modern large language models are built on the Transformer architecture~\cite{vaswani2017attention}, utilizing self-attention and feed-forward layers, along with other components, to compute the probabilities of output tokens autoregressively, i.e., conditioned on input and previously generated tokens. Self-attention and feed-forward layers can be implemented as a series of matrix multiplications and matrix-vector multiplications. Consequently, matrix multiplication dominates LLM inference runtime~\cite{karami2025perfhorizons}, making it a prime target for performance exploration. Additionally, researchers have developed optimizations like FlashAttention~\cite{dao2022flashattention}, an IO-aware implementation of self-attention that fuses two matrix multiplications and a softmax normalization to avoid repeated expensive round-trips to DRAM on devices like GPUs. %Given its importance in modern LLMs, we also explore the performance of FlashAttention in this work.

Modern LLM inference engines such as vLLM~\cite{kwon2023vllm} and llama.cpp~\cite{llama_cpp} provide the ability to interact with LLMs through chat interfaces and accelerate output generation by offloading the computation to GPUs. This process is split into two distinct phases, (1) \textit{prefill}, where the whole user prompt is fed into the model at once and the first output token is generated, and (2) \textit{decode}, where one new token is generated sequentially. The prefill stage can be computed using dense matrix multiplication, as all prompt tokens must be computed against all previous prompt tokens during self-attention, while the decode stage requires only one token to be computed given previous, cached, tokens using matrix-vector multiplication.

While server-based and multi-GPU engines like vLLM also incorporate sophisticated request-batching strategies in order to avoid the under-utilization of modern GPUs during decode phases, engines like llama.cpp are primarily built to run on a single user's machine and do not batch multiple requests. As WebGPU is designed for single-GPU browser-based applications, we focus on performance exploration of both matrix (for the prefill phase) and matrix-vector (for the decode phase) multiplication. Given its importance in modern LLM inference engines, we also explore the performance of dense FlashAttention, as it is run during the prefill phase.

\subsection{GPU Architecture}
GPUs are highly parallel, consisting of hundreds to thousands of cores organized hierarchically and executing in a Single-Instruction, Multiple-Thread (SIMT) manner \cite[\S 7.1]{cuda_guide_12_9}. Each core has its own set of registers and can perform basic computation. Small groups of cores, with sizes varying by GPU vendor and architecture, execute in tight synchronization (often in lockstep), allowing them to efficiently perform collective actions, e.g., shuffles and reductions. These small groups are organized into a compute unit, or a Streaming Multiprocessor (SM) in NVIDIA terminology, which share an L1 cache and can efficiently synchronize access to a special region of scratchpad memory. At the top level, these compute units also share access to an L2 cache to avoid frequent round-trips to main memory.

\subsection{WebGPU Programming Model}
Most GPU frameworks, including WebGPU, consist of a host API which executes on the CPU and communicates with the GPU driver, and a programming (or \textit{shading}) language which executes on the GPU. WebGPU's shading language (WGSL) provides an abstraction over lower-level GPU language and architecture concepts. While WebGPU is designed for both graphics and general computation, the focus of GPU Goldmines is on compute applications.

WGSL programs, called \textit{shaders} (which we refer to as \textit{kernels}, following common terminology in the machine learning communities), are written in a SIMT manner, with each invocation (thread) having access to unique built-in identifiers mapping them to the hierarchy of GPU cores. The smallest hierarchy of threads in WGSL is called a \textit{subgroup} (\textit{warp} in CUDA), which can perform efficient collective operations. Subgroups are organized into \textit{workgroups} (\textit{threadblocks} in CUDA), and can share and synchronize access to a scratchpad (called the \textit{workgroup} address space in WGSL). The total set of threads that execute a kernel is called the \textit{compute shader grid}, with the size of the set equal to the workgroup size multiplied by the number of workgroups.

%All threads in a grid have access to device-wide memory through input buffers which can either be in the \textit{storage} or \textit{uniform} address space. Storage buffers are general-purpose buffers that can be relatively large. These are either read-only or read-write, as opposed to the smaller and read-only uniform buffer. Additionally, uniform buffers can provide faster access times~\cite{}.

\paragraph{Subgroup Features} The WebGPU specification includes subgroup operations as an emerging feature, allowing threads within a subgroup to communicate directly without going through shared memory~\cite{webgpu-subgroup-matrix}. Of particular relevance to this work are two features: \textit{subgroupBroadcast}, which allows one thread to broadcast a value to all other threads in the subgroup, and \textit{subgroupMatrixMultiplyAccumulate}, which provides direct access to hardware matrix units such as tensor cores on RTX 5080, XMX engines on Arc B580, WMMA units on RX 7900XT, and specialized subgroup routines on M2. These features enable efficient matrix multiplication but are still maturing across WebGPU backends, with varying levels of support across browsers and devices. We leverage both features in our tunable kernel implementations (Sec.~\ref{sec:shaders}).

\subsection{WebGPU Implementations} There are several WebGPU implementations currently being developed: Dawn~\cite{dawn} for Chromium-based browsers, wgpu~\cite{wgpu} for Firefox, and a WebKit implementation~\cite{WebKit} for Safari. These implementations do not generate and run vendor-specific GPU programs directly; rather, they target several native backends depending on the operating system: DirectX~\cite{directx-specs} on Windows, Vulkan~\cite{vulkan1.3} on Linux and Android, and Metal~\cite{apple_metal} on macOS and iOS.

The variability introduced by multiple compiler implementations and backends underscores the need for research into performance portable WGSL kernels. Consequently, we discuss performance not only with respect to the GPU device, but also the \textit{context}, which includes the WebGPU implementation and intermediate framework being targeted, as these layers can substantially affect performance (Sec.~\ref{sec:shaders}).

\section{The GPU Goldmines Toolsuite}
\label{sec:framework}

We now present the GPU Goldmines toolsuite, which provides an interface for defining tunable kernels, a kernel miner that systematically explores the performance under different parameter spaces, and a centralized database that aggregates results across devices. By running entirely in the browser, GPU Goldmines can collect performance data from any WebGPU-capable device without requiring installation or platform-specific dependencies. An overview is shown in Figure~\ref{fig:overview}.

\subsection{Tunable Kernel Interface}

The toolsuite begins with the tunable kernel specification interface. This includes the templated kernel code, its arguments, tuning parameters, and auxiliary functions for performance measurements and output validation. It is designed to be flexible enough to support diverse tuning behaviors, yet structured enough to enable systematic data collection.

\begin{table}[t]
\centering
\small
\caption{Fields of a GPU Goldmines kernel argument.}
\label{tab:shader-args}
\begin{tabular}{lp{.76\columnwidth}}
\toprule
\textbf{Field} & \textbf{Description} \\
\midrule
%\rowcolor{lightgray}
\texttt{name} & Identifier matching the argument name in the kernel. \\
\texttt{type} & Data type (e.g., \texttt{u32}, \texttt{array<f16>}, \texttt{Q4\_0}). \\
%\rowcolor{lightgray}
\texttt{I/O} & Whether the argument is input, output, or both. \\
\texttt{region} & WebGPU memory region (\texttt{storage} or \texttt{uniform}). \\
%\rowcolor{lightgray}
\texttt{sizes} & Possible array sizes to enumerate over. \\
\texttt{fill} & Initialization strategy (e.g., zero-filled, random). \\
%\rowcolor{lightgray}
\texttt{values} & For scalar types, the possible values. \\
\bottomrule
\end{tabular}
\end{table}

\paragraph{Kernel Arguments} The interface requires specifying representative arguments to the tunable kernel. Each argument is defined by a name and a type (Table~\ref{tab:shader-args}). GPU Goldmines supports scalar types for signed, unsigned, and atomic integers, array types for integers, 32-bit, and 16-bit floating-point values with 2- and 4-tuple vector types, and quantized data formats, currently supporting q4\_0, q4\_1, and q6\_k, which are common formats found in llama.cpp.

Because GPU kernels return results by writing to memory, array arguments require additional specification. They may be classified as inputs, outputs, or both, and must specify their WebGPU memory region (\texttt{storage} or \texttt{uniform}) and initialization strategy. GPU Goldmines offers several initialization modes, including zero-filled and random floats. Scalar arguments are simpler, as they are input-only and do not require memory initialization.

Multiple input configurations can be specified using lists for sizes and values. For example, a matmul kernel might specify several matrix sizes of interest, and performance data will be collected for all specified sizes. When tuning for a more general library, it is useful to specify representative sizes, whereas when tuning for a specific AI model, the precise weight matrix sizes can be directly provided. The interface includes a \texttt{valid\_args} function that filters invalid argument combinations, e.g., requiring the inner matrix dimensions to match for matrix multiplication.

\paragraph{Tunable Kernels and Parameters} The interface includes a WGSL kernel code template, along with a host launch function \texttt{compute\_grid}, which computes the thread and workgroup layout for the kernel. Tuning parameter names can appear anywhere in this code and are denoted with double braces (e.g., \texttt{\{\{tile\_size\}\}}). The kernel arguments are required to have a one-to-one match to the names and types declared in the kernel argument specification.

Each tuning parameter is defined by a name and a list of possible values. The interface supports a \texttt{valid\_params} function that filters out invalid combinations, e.g., configurations that exceed device limits such as maximum workgroup size or shared memory capacity. These limits can be specified in the configuration or queried dynamically from the WebGPU adapter at runtime. The \texttt{compute\_grid} function is required in the specification because the thread layout can be impacted by input sizes and tuning parameters (e.g., the workgroup tile size).

\paragraph{Validation and Metrics} Two auxiliary functions can be included in the specification: \texttt{validate}, which checks kernel output arrays against a reference implementation, and \texttt{perf\_metric}, which computes kernel-specific performance metrics such as GFLOPs or achieved bandwidth.

\subsection{The Kernel Miner}

The kernel miner takes a kernel specification and explores tuning parameters and input arguments. The miner currently operates in three modes: exhaustive, heuristic, and warm-start heuristic, which can be summarized as follows:
\begin{itemize}
    \item \textbf{Exhaustive}: The miner enumerates all combinations of kernel arguments and tuning parameters via their Cartesian product. 
    \item \textbf{Heuristic}: The miner client communicates with a simple server running the mature OpenTuner framework (version 0.8.8)\footnote{\url{https://github.com/jansel/opentuner/tree/ac1408a}}. The client server architecture is used because OpenTuner cannot easily be ported to the browser.
    \item \textbf{Warm-start heuristic}: This is the same as the heuristic mode, with the difference being that this mode can provide a warm-start using values from previous runs stored in the performance database.
\end{itemize}

In each case, the miner filters values using the constraint functions \texttt{valid\_args} and \texttt{valid\_params} and the device limits. For each valid configuration, the kernel template is instantiated, GPU buffers and pipeline resources are allocated, and the kernel is executed on the target device in two phases: warmup and measurement. During warmup, the kernel is run 16 times to stabilize performance and outputs are validated against the reference implementation. During measurement, the kernel is executed 128 times and the average execution time is recorded along with kernel-specific performance metrics such as GFLOPs. We found that these warmup and iteration counts provide results with low variability.

\subsection{Performance Database}

All performance results from the kernel miner are stored in a centralized database on a backend server, indexed by device, browser, operating system, kernel, input dimensions, and tuning parameters. Because browsers can limit the amount of GPU information that can be queried for security reasons, we provide an option for the user to provide the name of their GPU. Otherwise, if the precise name cannot be determined, we use the name provided by \texttt{GPUAdapterInfo} on a best-effort basis. In future work, we plan on using performance clustering to infer the device being used, rather than relying on exact names. The database grows continuously as new devices visit the site and contribute results. We describe how this data is used to construct portable configurations and predict tuning parameters for new execution contexts in Sec.~\ref{sec:database}.

\section{Tunable LLM Inference Kernels}
\label{sec:shaders}
\label{sec:llama-shaders}

We implement tunable kernels for three performance-critical LLM inference routines: matrix multiplication (\textit{matmul}), matrix-vector multiplication (\textit{mvmul}), and flash attention (\textit{flash}). These routines span distinct performance profiles: \textit{matmul} is typically compute-bound and central to prefill, \textit{mvmul} is typically memory-bound and central to decode in single-batch systems such as llama.cpp, and \textit{flash} is a more complex fused kernel used in the prefill stage. Across different implementations and data signatures, we provide 13 tunable kernels (7 matmul, 5 mvmul, 1 flash). 

To our knowledge, these implementations are the state of the art for WebGPU, as the language does not yet support advanced features, like asynchronous memory transfers within the kernel or fine-grained synchronization required for split/stream-K optimizations~\cite{osama2023streamk}. 

\subsection{Matrix Multiplication Kernels}
\label{sec:matmul-shaders}

We implement several \textit{matmul} kernel families over input sizes $(M,N,K)$, where $A \in \mathbb{R}^{M \times K}$, $B \in \mathbb{R}^{K \times N}$, and $C \in \mathbb{R}^{M \times N}$. We support fp16 and fp32 (when available), as well as llama.cpp-style signatures in which activations are fp32 and weights are fp16 or quantized (q4\_0, q4\_1, q6\_k). For llama.cpp signatures, weights in right-hand matrices are stored transposed, which simplifies access patterns and enables direct integration; quantized values are expanded to fp16 values when loaded from global memory to local memory (following other llama.cpp kernels). Across implementations and data signatures, we provide 7 tunable \textit{matmul} kernels.

\paragraph{matmul-tiled}
is a conventional tiled matrix multiplication kernel using local memory, 2D thread tiling, and vectorized memory accesses. It requires no WebGPU extensions and is therefore the most portable implementation. The kernel exposes five tuning parameters: workgroup dimensions $(W_x,W_y)$, shared-memory K tile size $K_{\mathit{tile}}$, and thread-tile dimensions $(T_x,T_y)$. Each parameter ranges over powers of two, up to 128 for each workgroup dimension and $K_{\mathit{tile}}$, and up to 8 for each thread-tile dimension. Invalid configurations are filtered using device limits on workgroup size and shared memory. This yields 1677 valid tuning configurations for fp32 and 1737 for fp16.

\paragraph{matmul-bcst} is adapted from the matrix multiplication kernel used by ONNX Runtime Web. It uses the WebGPU subgroup broadcast operation to broadcast values of matrix $A$ within a subgroup, while matrix $B$ is staged in shared memory. This kernel requires the subgroup extension, which is supported on many (but not all) modern GPUs. It exposes four tuning parameters: workgroup dimensions, $K_{\mathit{tile}}$, and output rows per subgroup. The latter is restricted to powers of two up to 16. This yields 48 valid tuning configurations for fp16.

\paragraph{matmul-mma} is adapted from the matrix multiplication kernel used by llama.cpp and leverages hardware matrix units through WebGPU subgroup MMA operations. This requires the experimental MMA WebGPU extension, which is not widely supported at this time, but is actively being developed. These map to tensor cores on NVIDIA, XMX units on Intel, WMMA units on AMD, and specialized acceleration on Apple GPUs. The kernel tiles the output across subgroups and uses cooperative shared-memory loading. Since subgroup MMA dimensions are hardware-dependent, the kernel queries them at runtime; in our experiments, we found only the following dimensions are supported from WebGPU: Apple uses $8 \times 8$, AMD uses $16 \times 16$, and NVIDIA and Intel use $8 \times 16$. Relative to \textit{matmul-tiled}, this kernel replaces thread-tile parameters with subgroup fragment tile parameters $(S_x,S_y)$. Because WebGPU subgroup MMA currently supports only fp16 inputs, we implement this family only for fp16 (and narrower) weight formats. This yields 720 valid tuning configurations.

\subsection{Matrix-Vector Multiplication Kernels}
\label{sec:mvmul-shaders}

For matrix-vector multiplication, we implement a single tunable kernel family, \textit{mvmul}, over input sizes $(N,K)$, where $A \in \mathbb{R}^{N \times K}$, $x \in \mathbb{R}^{K}$, and $y \in \mathbb{R}^{N}$. We obtain this family by porting the CLBlast~\cite{nugteren2018clblast} OpenCL kernel (release 1.6.1)\footnote{\url{https://github.com/CNugteren/CLBlast/tree/1.6.1}} to WebGPU. The design uses vectorized loads, tiles matrix columns in shared memory, and performs a tree-based reduction across workgroup threads. We instantiate it for fp32, fp16, and llama.cpp-style signatures, including quantized weights (q4\_1, q6\_k). The kernel exposes four tuning parameters: workgroup dimensions, column tile size (powers of two up to 512), and rows per workgroup (powers of two up to 256), yielding 256 valid configurations for fp32 and 256 for fp16. Across data signatures, this gives a total of 5 \textit{mvmul} kernels.

\subsection{Flash Attention}
\label{sec:flash-attn}

We implement a tunable \textit{flash} kernel for attention over queries $Q \in \mathbb{R}^{M \times d}$, keys $K \in \mathbb{R}^{N \times d}$, and values $V \in \mathbb{R}^{N \times d}$, producing outputs $O \in \mathbb{R}^{M \times d}$. The kernel uses online softmax and subgroup MMA operations for both the query-key and attention-value matrix multiplications. We currently support the fp16 llama.cpp-style signature, in which queries and outputs are stored in fp32 while keys and values are stored in fp16. Shared memory is used for the query tile, intermediate attention scores, and the running row-wise statistics required by online softmax. The kernel also supports grouped-query attention (GQA), where multiple query heads share a single key-value head, as in the Llama3.2 and Qwen3 models used in our evaluation.

The kernel exposes three tuning parameters: the standard workgroup dimensions and KV tile size, which can be multiples of 8, starting at 8 and going up to size 64. This provides a total tuning space of 32 configurations. It additionally supports model-specific features including attention masking, attention sinks, and logit soft-capping. We leave systematic exploration of these additional configuration dimensions to future work.

\begin{table}[t]
\centering
\scriptsize
\caption{Summary of GPUs used in evaluation. Type: I = Integrated, D = Discrete. BW = Bandwidth (GB/s).}
\label{tab:gpus}
\begin{tabular}{lllcrr}
\toprule
\textbf{Vendor} & \textbf{GPU} & \textbf{OS} & \textbf{Type} & \textbf{BW} & \textbf{TFLOPS} \\
\midrule
\rowcolor{lightgray}
NVIDIA & RTX 5080   & Linux/Windows & D & 960  & 56.3 \\
% \rowcolor{lightgray}
       & RTX 4070   & Linux & D & 504  & 29.1 \\
\rowcolor{lightgray}
AMD    & RX 7900XT  & Linux   & D & 800  & 51.6 \\
% \rowcolor{lightgray}
Intel  & Arc B580   & Linux/Windows & D & 456  & 13.7 \\
\rowcolor{lightgray}
       & Arc A770   & Linux & D & 560  & 19.7 \\
Apple  & M2         & macOS   & I & 100  & 3.6  \\
\rowcolor{lightgray}
       & A17 Pro    & iOS     & I & --  & --  \\
% \rowcolor{lightgray}
Samsung & Xclipse 940 & Android & I & -- & -- \\
\bottomrule
\end{tabular}
\end{table}

% \begin{table}[t]
% \centering
% \scriptsize
% \caption{Summary of GPUs used in evaluation. Type: I = Integrated, D = Discrete. BW = Bandwidth (GB/s).}
% \label{tab:gpus}
% \rowcolors{1}{lightgray}{white}
% \renewcommand{\arraystretch}{1.15}
% \rowcolors{1}{lightgray}{white}

% \begin{tabular}{lllcrr}
% \toprule
% \rowcolor{white}
% \textbf{Vendor} & \textbf{GPU} & \textbf{OS} & \textbf{Type} & \textbf{BW} & \textbf{TFLOPS} \\
% \midrule
% \multirow[c]{2}{*}[-0.35ex]{NVIDIA} & RTX 5080    & Linux/Windows & D & 960 & 56.3 \\
%                                     & RTX 4070    & Linux         & D & 504 & 29.1 \\
% AMD                                 & RX 7900XT   & Linux         & D & 800 & 51.6 \\
% \multirow[c]{2}{*}[-0.35ex]{Intel}  & Arc B580    & Linux/Windows & D & 456 & 13.7 \\
%                                     & Arc A770    & Linux         & D & 560 & 19.7 \\
% \multirow[c]{2}{*}[-0.35ex]{Apple}  & M2          & macOS         & I & 100 & 3.6  \\
%                                     & A17 Pro     & iOS           & I & XX  & XX   \\
% Samsung                             & Xclipse 940 & Android       & I & XX  & XX   \\
% \bottomrule
% \end{tabular}
% \end{table}

\begin{table}[t]
\centering
\caption{Input sizes derived from two LLM models and used to evaluate \textit{matmul} ($M>1$) and \textit{mvmul} ($M=1$). Blue cells denote pairs evaluated only for \textit{matmul}, green cells denote pairs evaluated only for \textit{mvmul}.}
\label{tab:sizes}
\begin{tabular}{cll}
\hline
\textbf{M} & \textbf{Llama3.2 (N$\times$K)} & \textbf{Qwen3 (N$\times$K)} \\
\hline
\multirow[c]{6}{*}{$\{1,32,64,4096\}$} & \cellcolor{matmulonly}512$\times$2048   & \cellcolor{matmulonly}1024$\times$1024 \\
                         & \cellcolor{matmulonly}2048$\times$2048  & \cellcolor{matmulonly}1024$\times$2048 \\
                         &                          & \cellcolor{matmulonly}2048$\times$1024 \\
                         & 8192$\times$2048                         & 1024$\times$3072 \\
                         & 2048$\times$8192                         & 3072$\times$1024 \\
                         & \cellcolor{mvmulonly}128256$\times$2048 & \cellcolor{mvmulonly}151936$\times$1024 \\
\hline
\end{tabular}
\end{table}

\section{Experimental Setup}
\label{sec:setup}
We evaluate our tunable kernels across 8 GPUs from 5 vendors (Table~\ref{tab:gpus}), covering discrete and integrated GPUs spanning NVIDIA, AMD, Intel, Apple, and Samsung. Some of these devices have similar architectures (e.g., from the same vendor), while others are extremely different (e.g., mobile and discrete devices). We note that different OSes will provide different WebGPU backends (Vulkan for Linux and DirectX for Windows), and thus different performance profiles. 

We select matrix sizes obtained from benchmarking Llama3.2 1B and Qwen3 0.6B in llama.cpp. We vary $M$, where $M = 1$ corresponds to \textit{mvmul} kernels used in decode; $M = 32,64$ correspond to prefill sizes for short contexts, e.g., that would be used in chat, and $M=4096$ corresponds to a large prefill, e.g., that would be used in document summary. These sizes are summarized in Table~\ref{tab:sizes}. Although single batch decode is conceptually equivalent to matrix multiplication with $M=1$, framework-level operator specialization and partitioning can lead to different effective $N\times K$ dimensions in practice. For example, Llama3.2 1B uses 32 attention heads with a head dimension of 64, while Qwen3 0.6B uses 16 heads of dimension 128, so we tune over these sizes.

We select four GPUs (RTX 5080, RX 7900XT, Arc B580, and M2) to exhaustively explore tuning parameters across a subset of kernels, sizes and execution contexts. This allows us to characterize the performance portability across various dimensions while keeping exploration time feasible (described in Sec.~\ref{sec:perf}). The remaining four GPUs (RTX 4070, Arc A770, A17 Pro, and Xclipse 940) are benchmarked using more sparse approaches to evaluate how well our approaches adapt to previously unseen devices (Sec.~\ref{sec:warm-start}).

\begin{figure*}[t]
    \centering

    \begin{subfigure}{0.32\textwidth}
        \centering
        \includegraphics[width=\linewidth]{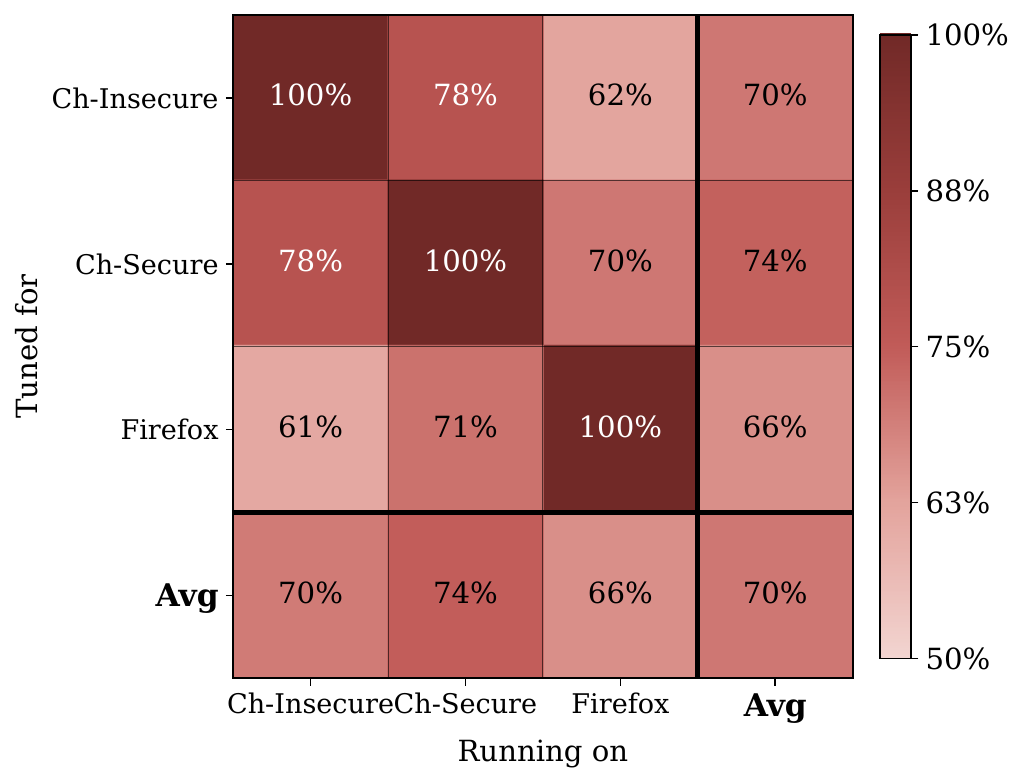}
        \caption{Cross-browser portability on RTX 5080.}
        \label{fig:perfport-crossbrowser}
    \end{subfigure}
    \hfill
    \begin{subfigure}{0.32\textwidth}
        \centering
        \includegraphics[width=\linewidth]{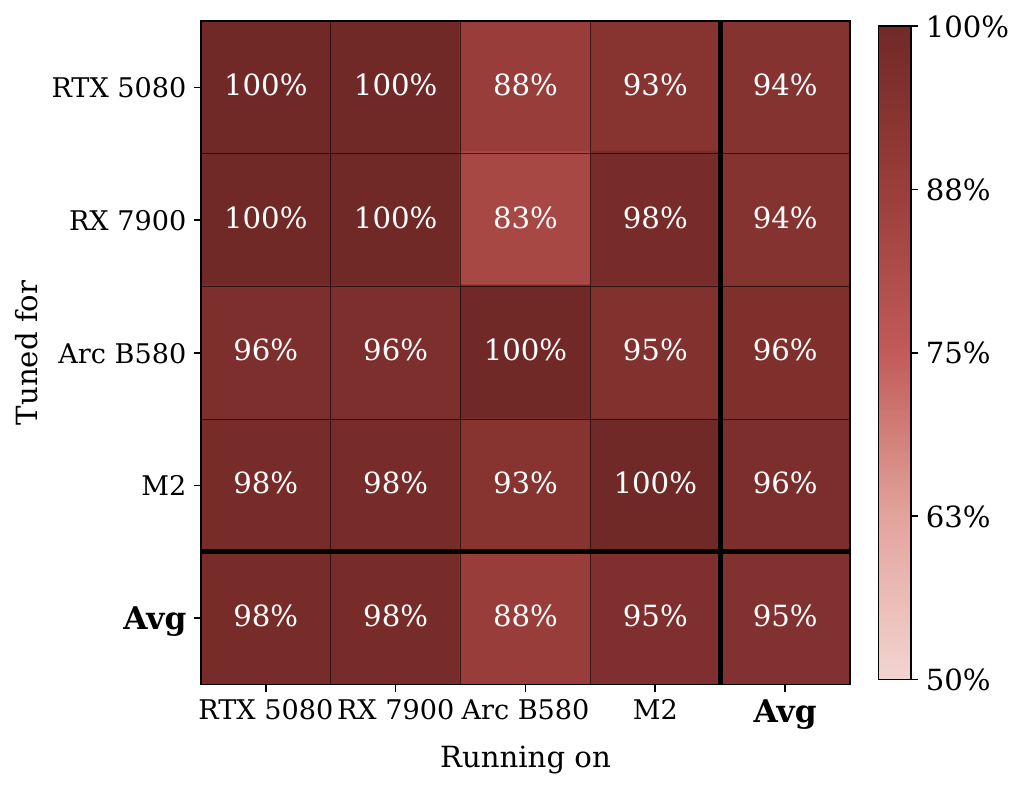}
        \caption{\textit{mvmul} portability across 4 devices.}
        \label{fig:perfport-mvmul}
    \end{subfigure}
    \hfill
    \begin{subfigure}{0.32\textwidth}
        \centering
        \includegraphics[width=\linewidth]{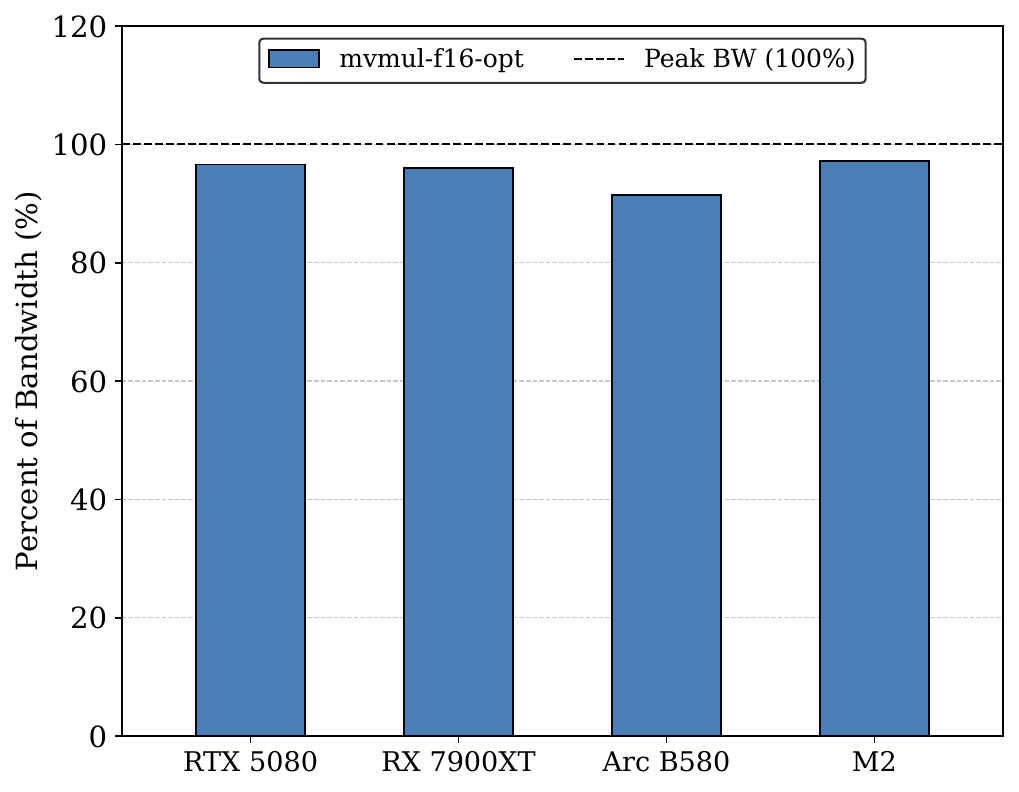}
        \caption{Peak bandwidth of \textit{mvmul} across devices.}
        \label{fig:perfport-mvmul-bw}
    \end{subfigure}

    \caption{Performance portability of matrix multiplication and matrix-vector multiplication in different environments.}
    \label{fig:main}
\end{figure*}

\section{Characterizing Performance Portability}
\label{sec:perf}

In this section, we use GPU Goldmines' exhaustive exploration across a subset of our kernels to characterize performance portability across several dimensions. Our main result is that \textit{matmul} is substantially more sensitive to tuning parameters than \textit{mvmul} or \textit{flash}, motivating our later focus on \textit{matmul}. We also show that smaller input sizes are less portable and have less predictable optimal tuning parameters.

\subsection{Devices and Browser Environments}

We exhaustively evaluate all tuning parameter combinations for the fp16 \textit{matmul-tiled} and \textit{mvmul} kernels across the 27 LLM sizes described in Table~\ref{tab:sizes}, as well as the 6 \textit{flash} sizes. For a source device $d$, target device $d'$, and input $i$, we define retained performance as the performance achieved on $d'$ using the best tuning parameters from $d$, normalized by the optimal performance observed on $d'$ for $i$. We then average across inputs.

\paragraph{Matrix Multiplication}
The cross-device portability of \textit{matmul-tiled} was shown in Figure~\ref{fig:perfport}. On average, a configuration tuned for one device retains 77\% of optimal performance when transferred to another, with individual cases dropping as low as 40\% (e.g., M2's preferred small-workgroup configuration on RTX 5080 at $64\times2048\times2048$). RX 7900XT provides the most portable source configurations, achieving 82\% of optimal performance on average across other devices, while M2 provides the least portable, achieving only 69\%. Viewed from the target side, RX 7900XT is also the most robust device, retaining an average of 83\% performance when executing configurations tuned elsewhere, while Arc B580 is the least robust, with M2 only 1\% better on average.

Performance varies across browser environments, even on the same device. We therefore evaluate the same \textit{matmul-tiled} kernel on the RTX 5080 under three browser configurations: Chrome without bounds checking (Ch-insecure), Chrome with bounds checking (Ch-secure), and Firefox. Kernels tuned for one browser configuration retain only 70\% of optimal performance on average when run in another (Figure~\ref{fig:perfport-crossbrowser}). Cross-browser performance differs by up to 39\%, with the widest gap between Firefox and Chrome without bounds checking, where configurations tuned for one retain only 61\% of optimal performance on the other. Changing only the Chrome security setting reduces performance by 26\% on average in one direction and 30\% in the other. Thus, tuning for a device alone is not sufficient; the browser and its configuration also affect which tuning parameters perform best.

For the remainder of this study, we run experiments only on Chrome with security checks enabled, as this does not affect the generality of our approach while keeping the experimental space manageable. When we refer to different devices below, we do so in the sense of their overall execution environment, where the device includes the browser and other relevant performance-affecting configuration choices.

\paragraph{Matrix-Vector Multiplication and Flash Attention}

We perform the same cross-device study for the fp16 \textit{mvmul} kernel, with results shown in Figure~\ref{fig:perfport-mvmul}. In contrast to \textit{matmul}, \textit{mvmul} is much less sensitive to tuning parameters; transferred configurations retain 95\% of optimal performance on average, with a worst case of 83\%. This is likely because \textit{mvmul} is memory-bandwidth-bound, so once data is loaded efficiently, the exact parameter values matter less. Consistent with this, these kernels already reach at least 80\% of peak bandwidth across devices (Figure~\ref{fig:perfport-mvmul-bw}), leaving limited room for additional tuning improvements.\footnote{We do not make a similar comparison for \textit{matmul}, since current WebGPU limitations, such as the lack of asynchronous memory copies, prevent many devices from approaching their hardware limits.} We therefore do not pursue \textit{mvmul} further, nor do we emphasize LLM decode performance later in the paper, since \textit{mvmul} is its primary kernel.

We also perform the cross-device study for \textit{flash}. The results show that transferred configurations retain 89\% of optimal performance on average. This may partly reflect the smaller tuning space we expose for \textit{flash}, which is harder to parameterize systematically given the complexity of the kernel. Combined with the fact that \textit{flash} is more difficult to benchmark accurately because of its sensitivity to context length, this result led us to focus the remainder of the paper on \textit{matmul}.

\begin{figure}[t]
\centering
\includegraphics[width=\columnwidth]{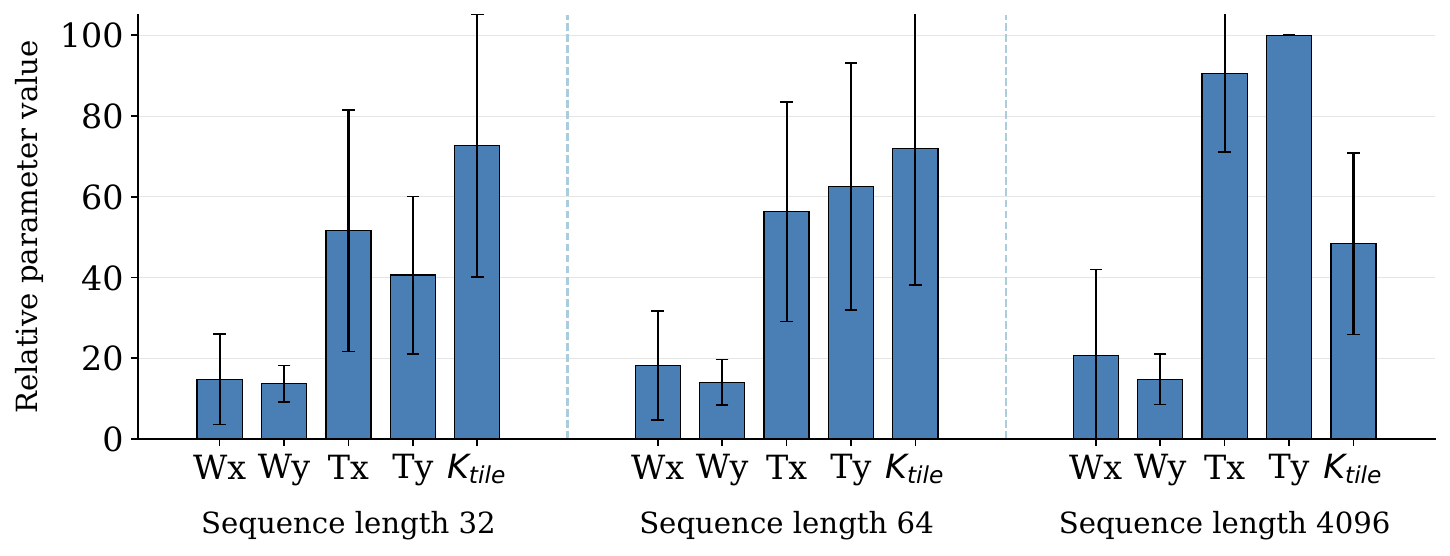}
\caption{Average optimal tuning parameter values for different sequence lengths, normalized to each parameter's maximum, along with their variance across devices and $(N,K)$ sizes.}
\label{fig:tuningparams}
\end{figure}

\subsection{Portability Across Matrix Multiplication Input Sizes}
\label{sec:portability}
Having established \textit{matmul} as the most tuning-sensitive kernel, we now study its portability across input sizes, focusing on sequence length. We consider two small values ($M=32,64$) and one large value ($M=4096$). Breaking down the device-portability heatmap of Figure~\ref{fig:perfport} by sequence length, we find that portability at $M=4096$ is relatively high, at 89\%, whereas portability at $M=32$ and $M=64$ drops to 72\% and 68\%, respectively. Larger matrix multiplications are therefore more stable across devices, while smaller ones are substantially less stable.

Figure~\ref{fig:tuningparams} helps explain this effect by showing the optimal tuning parameters and their variance for different sequence lengths. Large sizes consistently favor larger $K$ tiles and larger register tiling dimensions, with relatively little variance across devices and $(N,K)$ pairs. In contrast, the smaller sizes exhibit much higher variance, indicating that their optimal configurations are more architecture dependent. This is especially important for edge LLMs, since these smaller sizes arise frequently in chat-oriented workloads.

\section{Utilizing the Performance Database}
\label{sec:database}

We next evaluate how GPU Goldmines uses previously collected performance data to select tuning parameters and accelerate tuning, focusing on the fp16 \textit{matmul-tiled} kernel. We consider three settings of increasing difficulty: selecting portable configurations for known input sizes using data aggregated across devices, warm-starting search for new devices, and transferring configurations to unseen sizes using a distance metric over tensor dimensions. Throughout this section, we assume the database is seeded with the exhaustive runs from the previous section.

\begin{table}[t]
\centering
\small
\caption{Average retained performance of portable selection strategies and a cross-device transfer baseline, grouped by sequence length $M$.}
\label{tab:portable}
\begin{tabular}{lccc}
\toprule
\textbf{Sequence Length} & \textbf{Transfer} & \textbf{Min-Max} & \textbf{Min-Avg} \\
\midrule
$M = 32$   & 72\% & 83\% & 87\% \\
$M = 64$   & 68\% & 83\% & 86\% \\
$M = 4096$ & 89\% & 96\% & 97\% \\
\midrule
\textbf{Average} & 77\% & 88\% & 90\% \\
\bottomrule
\end{tabular}
\end{table}

\subsection{Portable Configuration Selection for Known Sizes}
\label{sec:portable-select}

We first consider the case where the input size is known and represented in the database, and the goal is to select a single configuration that works well across devices for that size. This setting is useful for libraries that do not perform dynamic device checks at runtime. We evaluate \textit{matmul-tiled} on our four exhaustive GPUs using retained performance, which requires exhaustive data so that the device-optimal performance for each size is known.

For each input size, we identify two portable configurations from the database:
\begin{itemize}
    \item \textit{Min-Avg} selects the tuning parameters that maximize average retained performance across devices.
    \item \textit{Min-Max} selects the tuning parameters that maximize minimum retained performance across devices.
\end{itemize}

Table~\ref{tab:portable} summarizes the results. Min-Avg achieves 90\% retained performance on average, while Min-Max achieves 88\%. Both improve over the transfer baseline (77\%), where each device's best configuration is transferred to other devices. Min-Max raises the worst-case floor from 40\% under transfer to 70\%, while Min-Avg achieves a higher average with a worst case of 60\% in our fine-grained results. On 6 of 12 sizes, both strategies select the same configuration, indicating no tradeoff between average and worst-case performance at those sizes. At $M=4096$, both strategies achieve more than 95\%, consistent with the previous section's finding that larger sequence lengths are more stable across devices.

\subsection{Warm-Start Tuning for Unknown Devices}
\label{sec:warm-start}
We next consider the case where the input sizes are present in the database, but the device is not. We warm-start autotuning with six initial seeds: the best configuration from each of the four known devices, together with the Min-Max and Min-Avg portable configurations above. We integrate this approach with OpenTuner~\cite{ansel2014opentuner} and evaluate it on 8 Llama 3.2 1B \textit{matmul} sizes at $M=32$ and $M=64$, where the portability gap is largest (Table~\ref{tab:portable}).

We compare seeded and unseeded autotuning on five devices: three known (RTX 5080, RX 7900XT, Arc B580) and two unknown (Apple M1, Qualcomm X1-85), running 10 independent trials of 50 iterations each (Table~\ref{tab:seeded}). Seeded search converges 52\% faster on average (iteration 14 vs 29) and finds configurations with 4.9\% higher performance. The effect is strongest on known devices, but also holds on the two unknown devices, indicating that portable configurations transfer useful structure to unseen hardware.

\begin{table}[t]
\centering
\footnotesize
\caption{Seeded vs.\ non-seeded autotuning on known (*) and unknown devices. Results are averaged over 10 trials on 8 Llama 3.2 1B \textit{matmul} sizes ($M = 32, 64$). Best iteration is reported as mean [CI lo, CI hi]. Perf+ gives average seeded performance relative to unseeded.}
\label{tab:seeded}
% \begin{tabular}{lccr}
% \toprule
% \textbf{GPU} & \textbf{Seeded} & \textbf{Unseeded} & \textbf{Perf+} \\
% \midrule
% 5080*   & 5.1  [3.7, 6.5] & 31.8 [29.1, 34.4] & +4.3\% \\
% 7900XT* & 2.7  [2.1, 3.3] & 26.9 [24.1, 29.8] & +10.6\% \\
% B580*   & 18.5 [15.6, 21.3] & 30.4 [27.7, 33.2] & +3.7\% \\
% M1      & 22.7 [20.0, 25.5] & 30.5 [27.6, 33.4] & +5.1\% \\
% X1-85   & 21.5 [18.7, 24.3] & 27.3 [24.6, 30.0] & +0.7\% \\
% \midrule
% \textbf{Avg.} & 14.0 & 29.4 & \textbf{+4.9\%} \\
% \bottomrule
% \end{tabular}
\begin{tabular}{
    l
    S[table-format=2.1]
    l
    S[table-format=2.1]
    l
    r
}
\toprule
\textbf{GPU} & \multicolumn{2}{c}{\textbf{Seeded}} & \multicolumn{2}{c}{\textbf{Unseeded}} & \textbf{Perf+} \\
\midrule
5080*   &  5.1 & [3.7, 6.5]    & 31.8 & [29.1, 34.4] & +4.3\% \\
7900XT* &  2.7 & [2.1, 3.3]    & 26.9 & [24.1, 29.8] & +10.6\% \\
B580*   & 18.5 & [15.6, 21.3]  & 30.4 & [27.7, 33.2] & +3.7\% \\
M1      & 22.7 & [20.0, 25.5]  & 30.5 & [27.6, 33.4] & +5.1\% \\
X1-85   & 21.5 & [18.7, 24.3]  & 27.3 & [24.6, 30.0] & +0.7\% \\
\midrule
\textbf{Avg.} & \multicolumn{2}{c}{14.0} & \multicolumn{2}{c}{29.4} & \textbf{+4.9\%} \\
\bottomrule
\end{tabular}
\end{table}

\subsection{Parameter Selection for Unknown Sizes}

We finally consider the case where the device is represented in the database but the input size is not. Rather than exhaustively re-tuning each new size, we predict a portable configuration from nearby sizes. We use the 12 Llama 3.2 1B sizes already in the database and evaluate on 15 unseen Qwen3 0.6B sizes, which share the same sequence lengths $M$ but differ in $N$ and $K$ (Table~\ref{tab:sizes}).

Guided by a series of smaller interactive pilot experiments, we identified the following method: For a new size $s_{\text{new}}=(M,N,K)$, we select the nearest training size in Euclidean $\log_2$ space and reuse its Min-Avg portable configuration to avoid overfitting. Using $\log_2$ makes similarity depend on relative scale rather than absolute difference, which better matches how GPU tuning behavior changes across problem sizes. We use Min-Avg rather than a device-specific optimum to avoid overfitting to the training devices.

Across the four exhaustive GPUs, this nearest-size rule achieves retained performance of 80\%, 77\%, and 96\% for $M=32,64$, and 4096, respectively, with an overall average of 84\%. Thus, even a simple distance-based rule transfers well to unseen matrix sizes, achieving 84\% retained performance, matching a Bayesian portable method~\cite{nadaraya1964estimating} we also explored that predicts configurations using distance-weighted scores from nearby sizes.

\section{Kernel and End-to-end Evaluation}
\label{sec:e2e}
We characterize how our tuned kernels perform relative to existing WebGPU implementations by comparing them against ONNX Runtime (commit \texttt{50f3cc3})\footnote{\url{https://github.com/microsoft/onnxruntime/tree/50f3cc3}}, SurfGrad (commit \texttt{4f696e3})\footnote{\url{https://github.com/zanussbaum/surfgrad/tree/4f696e3}}, and llama.cpp's WebGPU kernels (commit \texttt{6de97b9})\footnote{\url{https://github.com/ggml-org/llama.cpp/tree/6de97b9}}.

\begin{figure*}[t]
\centering
\includegraphics[width=\textwidth]{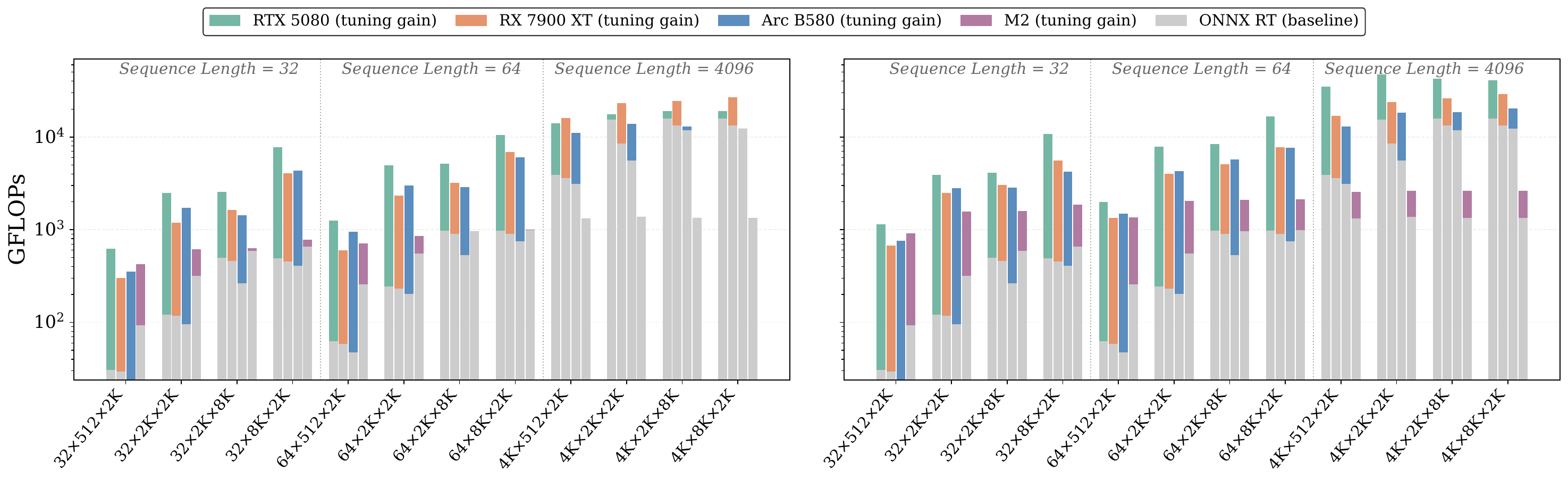}
\caption{GFLOPs of our two FP16 matmul variants across 12 Llama 3.2 1B inference sizes on 4 GPUs. Each bar shows the ONNX Runtime baseline (grey) and the additional performance from per-device tuning (colored). Left: matmul-bcst. Right: matmul-mma.}
\label{fig:onnx-comp}
\end{figure*}

\subsection{ONNX Runtime Comparison}
\label{sec:onnx-e2e}
\paragraph{ORT kernel description} ONNX Runtime (ORT) generates WebGPU kernels for matrix multiplication at model compilation time. We extracted and analyzed the generated kernels across all four devices and found that ORT produces a single fixed kernel for RTX 5080, RX 7900XT, and M2, with no per-device or per-size tuning. This kernel is essentially an instantiation of \textit{matmul-tiled} with a workgroup size of $8 \times 8$, a K tile of 32, and a thread tile size of 4 x 4. For Arc B580, ORT generates a second variant for $M \geq 64$ sizes that uses \textit{subgroupBroadcast} to distribute matrix $A$ values across subgroup lanes, corresponding to an instantiation of our \textit{matmul-bcst} kernel. This is the only specialization in ORT that we were able to identify.

\paragraph{Tuning the ORT kernel} We observed that the \textit{matmul-bcst} kernel ORT uses only for Intel can benefit all devices when properly tuned. By identifying and tuning four parameters (workgroup dimensions, output rows per subgroup, and K-dimension tile size), \textit{matmul-bcst} achieves a $3.7\times$ geometric mean improvement over ORT across all four devices and 12 input sizes. This suggests that even the same underlying kernel significantly benefits from per-device tuning rather than a fixed configuration.

\paragraph{matmul-mma vs ORT} When hardware matrix units are available, a tuned matmul-mma kernel shows a $7.0\times$ geometric mean improvement across all device-size combinations (Figure~\ref{fig:onnx-comp}, right panel). This is not surprising as these kernels take advantage of very efficient hardware acceleration. However, incorporating these kernels into a framework is presently difficult, as the MMA extensions are not widely supported and devices support different MMA dimensions.

\paragraph{Overall Best matmul Kernel} We do not show the results for \textit{matmul-tiled} here for brevity; however it can achieve the overall highest performance at smaller sizes on RTX 5080 and RX 7900XT (e.g., all $M=32$ and $M=64$ sizes on RX 7900XT). Thus, ultimately, a framework needs to not only identify high-performing tuning parameters, but also potentially switch kernel implementations; this is done in other libraries (e.g., cuBLAS~\cite{nvidia_cublas_2025}).

\subsection{SurfGrad Comparison}

The only stand-alone matrix multiplication library for WebGPU that we are aware of is SurfGrad~\cite{surfgrad}. It includes optimizations like thread tiling and manual loop unrolling. We compare our tuned \textit{matmul-tiled} implementation against SurfGrad and evaluate across the GPUs on three square matrix sizes: 1K, 2K, and 4K (the only supported shapes in SurfGrad). With per-device tuning, matmul-f32-opt shows a $6.5\times$ geometric mean improvement over SurfGrad across all device-size combinations. On discrete GPUs the tuned kernel consistently outperforms SurfGrad, while on M2 the improvement is $2.4\times$. 

%(Figure~\ref{fig:surfgrad-comp}).
\subsection{llama.cpp: Kernels and End to End}
\label{sec:llama-e2e}

Llama.cpp uses a single set of tuning parameters for its WebGPU Q4\_0 matmul kernel across all devices. We exhaustively explored the same kernel's tuning space using GPU Goldmines and identified per-device optimal configurations. Across all four devices and 12 Llama 3.2 1B sizes, the per-device best configurations show a $1.41\times$ geometric mean improvement over the llama.cpp defaults, with per-device averages ranging from $1.13\times$ on M2 to $1.80\times$ on RX 7900XT.

Since llama.cpp uses a single configuration for all matmul invocations during inference, we select one configuration per device by applying the Min-Avg criterion (Sec.~\ref{sec:portable-select}) across sizes rather than across devices: for each device, we choose the configuration with the highest average retained performance over all 12 inference sizes. We run llama.cpp through wllama (commit \texttt{0d389c7})\footnote{\url{https://github.com/reeselevine/wllama/tree/0d389c7}}, which builds llama.cpp at commit \texttt{6de97b9}. We evaluate end-to-end prefill performance using 100 prompts randomly sampled from MMLU~\cite{hendrycks2021mmlu} on Llama 3.2 1B~\cite{grattafiori2024llama3} and Qwen3 0.6B~\cite{yang2025qwen3} across all four devices. We observe a $1.48\times$ geometric mean prefill improvement. Decode performance is unchanged, as the tuning targets matmul (used in prefill) rather than mvmul (used in decode).

\section{Related Work}
\label{sec:related}

\paragraph{Performance Portability of GPU Kernels}
%Pennycook et al.~\cite{pennycook2016metricperformanceportability} formalize performance portability as the harmonic mean of efficiency across platforms, providing a quantitative framework for measuring portability gaps, but applies it to previous studies to measure portability instead of designing a toolsuite to maximize it like this work. GPTune~\cite{liu2021gptune} addresses performance portability through multitask Bayesian optimization that transfers tuning knowledge across problem sizes and machines, and its crowdsourced extension~\cite{cho2023gptunecrowd} accumulates community-contributed data for reuse on new platforms. Among portable GPU libraries, CLBlast~\cite{nugteren2018clblast} exposes 14 tuning parameters per OpenCL BLAS kernel and maintains a community-contributed database across approximately 50 devices. GPU Goldmines shares GPTune's goal of cross-context transfer and CLBlast's philosophy of parameterized kernels with a centralized database, but targets WebGPU's uniquely broad deployment space where browser implementations and compilers introduce variability beyond the hardware itself.

Pennycook et al.~\cite{pennycook2016metricperformanceportability} formalize performance portability as the harmonic mean of efficiency across platforms, providing a quantitative framework for measuring portability gaps. GPTune~\cite{liu2021gptune} addresses performance portability through multitask Bayesian optimization that transfers tuning knowledge across problem sizes and machines, and its crowdsourced extension~\cite{cho2023gptunecrowd} accumulates community-contributed data for reuse on new platforms. Among portable GPU libraries, CLBlast~\cite{nugteren2018clblast} exposes 14 tuning parameters per OpenCL BLAS kernel and maintains a community-contributed database across approximately 50 devices. GPU Goldmines shares GPTune's goal of cross-context transfer and CLBlast's philosophy of parameterized kernels with a centralized database, but targets WebGPU's uniquely broad deployment space where browser implementations and compilers introduce variability beyond the hardware itself.

\paragraph{LLM Inference on Consumer Hardware}
LLM in a Flash~\cite{alizadeh2024llminflash} enables large model inference under limited memory through selective weight loading from flash storage. FlexGen~\cite{sheng2023flexgen} achieves high-throughput inference on a single GPU through offloading to CPU and disk. PowerInfer~\cite{song2024powerinfer} exploits activation sparsity to partition computation between CPU and GPU on consumer devices, and PowerInfer-2~\cite{xue2024powerinfer2} extends this to smartphones. AWQ~\cite{lin2025awq} introduces activation-aware weight quantization that enables deployment of large models on mobile GPUs. Xu et al.~\cite{xu2024npu} achieve over 1000 tokens per second for on-device prefilling by offloading computation to NPUs. These systems address memory, quantization, and hardware acceleration challenges for edge deployment, but do not address how kernel tuning parameters transfer across diverse GPU architectures for LLM workloads.

\paragraph{WebGPU Kernel Optimization}
%Wang et al.~\cite{wang2024anatomizing} provide the first comprehensive measurement of in-browser deep learning inference, revealing up to 19.4$\times$ performance variation across devices due to hardware-kernel affinity. Several systems compile or generate WebGPU kernels through TVM-based pipelines: WebLLM~\cite{ruan2024webllm} compiles LLM models ahead of time, and nnJIT~\cite{jia2024nnjit} JIT-compiles kernels at runtime, reducing the search space to approximately 41 candidates through web-specific heuristics and sharing results via a crowdsourced database indexed by device ID. nnJIT shares a goal of tuning WebGPU kernels and even optionally provides the ability to pull kernels from a database of similar devices at runtime by clustering using microbenchmarks. However, their auto-tuning runs online, they do not investigate more advanced kernels that use subgroup matrices, and they do not rigorously explore performance portability beyond their simple clustering algorithm. In contrast, this paper introduces a systematic method for tuning and a rigorous toolsuite, along with decision procedures that can be utilized by anyone looking for more performance from their WebGPU kernels.

Wang et al.~\cite{wang2024anatomizing} provide the first comprehensive measurement of in-browser deep learning inference, revealing up to 19.4$\times$ performance variation across devices due to hardware-kernel affinity. Several systems compile or generate WebGPU kernels through TVM-based pipelines: WebLLM~\cite{ruan2024webllm} compiles LLM models ahead of time, and nnJIT~\cite{jia2024nnjit} JIT-compiles kernels at runtime, reducing the search space to approximately 41 candidates through web-specific heuristics and sharing results via a crowdsourced database indexed by device ID. While nnJIT shares GPU Goldmines' goal of tuning WebGPU kernels and reducing the search space, it mainly focuses on online auto-tuning one version of matrix multiplication, with an optional optimization that attempts to choose parameters from a database by clustering devices based on a small performance microbenchmark. In contrast, GPU Goldmines scales to systematically explore performance on many kernels through exhaustive tuning and auto-tuning, and provides a toolsuite with more sophisticated performance portability techniques. WeInfer~\cite{chen2025weinfer} optimizes the WebGPU runtime layer, improving buffer management and asynchronous scheduling, but targets the runtime rather than kernel implementations. GPU Goldmines operates at the kernel tuning level, systematically exploring how parameter choices affect performance and building a reusable database for portability-aware configuration selection.

%WeInfer~\cite{chen2025weinfer} optimizes the WebGPU runtime layer, improving buffer management and asynchronous scheduling. However, these compiler-generated kernels are not tuned per device, nnJIT's database shares per-device results rather than studying portability across device families, and WeInfer targets the runtime layer rather than kernel implementations. GPU Goldmines operates at the kernel tuning level, systematically exploring how parameter choices affect performance and building a reusable database for portability-aware configuration selection.

\section{Conclusion}

We presented GPU Goldmines, a browser-based framework that accumulates WebGPU kernel performance data across devices, browsers, and workload shapes. Matrix multiplication is the most tuning-sensitive LLM kernel, with smaller chat-oriented sizes being the least portable. From the database, we derive portable configurations recovering 90\% of device-optimal performance, size prediction achieving 84\% on unseen sizes, and seeded autotuning converging 52.4\% faster, yielding $1.48\times$ prefill improvement in llama.cpp. As WebGPU support broadens across devices and backends, GPU Goldmines' extensible design allows the database and kernel families to grow alongside the WebGPU ecosystem.
\footnote{The authors used ChatGPT and Claude as editorial aids. All concepts, technical claims, organization, and conclusions were determined by the authors and carefully validated.}

\bibliographystyle{ACM-Reference-Format}
\bibliography{refs}

@inproceedings{price2017perfport,
author = {Price, James and McIntosh-Smith, Simon},
title = {Analyzing and improving performance portability of OpenCL applications via auto-tuning},
year = {2017},
url = {https://doi.org/10.1145/3078155.3078173},
booktitle = {Proceedings of the 5th International Workshop on OpenCL},
series = {IWOCL '17}
}

@misc{webgpu,
  author       = {W3C},
  title        = {Web{GPU}},
  howpublished = {\url{https://www.w3.org/TR/webgpu/}},
}

@misc{vulkan1.3,
  author       = {Khronos Group},
  title        = {Vulkan 1.3 Specification},
  howpublished = {\url{https://registry.khronos.org/vulkan/specs/1.3-extensions/html/vkspec.html}},
}

@misc{apple_metal,
  author       = {{Apple Inc.}},
  title        = {Metal},
  howpublished = {\url{https://developer.apple.com/documentation/metal/}},
}

@misc{directx-specs,
  author       = {Microsoft},
  title        = {{DirectX} Specifications},
  howpublished = {\url{https://microsoft.github.io/DirectX-Specs/}},
}

@misc{dawn,
  title = {Dawn: A {WebGPU} Implementation},
  howpublished = {\url{https://dawn.googlesource.com/dawn}},
  year = {2024}
}

@misc{wgpu,
  author       = {Rust Graphics Mages},
  title        = {{w}gpu},
  howpublished = {\url{https://github.com/gfx-rs/wgpu}},
}

@misc{onnx,
  author       = {{Open Neural Network Exchange}},
  title        = {{O}NNX},
  howpublished = {\url{https://github.com/onnx}},
}

@misc{surfgrad,
  author       = {Nussbaum, Zach},
  title        = {{s}urfgrad},
  howpublished = {\url{https://github.com/zanussbaum/surfgrad}},
  year = {2025}
}

@inproceedings{vaswani2017attention,
author = {Vaswani, Ashish and Shazeer, Noam and Parmar, Niki and Uszkoreit, Jakob and Jones, Llion and Gomez, Aidan N. and Kaiser, \L{}ukasz and Polosukhin, Illia},
title = {Attention is all you need},
year = {2017},
booktitle = {Proceedings of the 31st International Conference on Neural Information Processing Systems},
series = {NIPS'17},
url = {https://doi.org/10.48550/arXiv.1706.03762}
}

@inproceedings{kwon2023vllm,
author = {Kwon, Woosuk and Li, Zhuohan and Zhuang, Siyuan and Sheng, Ying and Zheng, Lianmin and Yu, Cody Hao and Gonzalez, Joseph and Zhang, Hao and Stoica, Ion},
title = {Efficient Memory Management for Large Language Model Serving with PagedAttention},
year = {2023},
url = {https://doi.org/10.1145/3600006.3613165},
booktitle = {Proceedings of the 29th Symposium on Operating Systems Principles},
series = {SOSP '23}
}

@inproceedings{dao2022flashattention,
author = {Dao, Tri and Fu, Daniel Y. and Ermon, Stefano and Rudra, Atri and R\'{e}, Christopher},
title = {FLASHATTENTION: fast and memory-efficient exact attention with IO-awareness},
year = {2022},
booktitle = {Proceedings of the 36th International Conference on Neural Information Processing Systems},
series = {NIPS '22},
url = {https://doi.org/10.48550/arXiv.2205.14135}
}

@article{zheng2025edge,
author = {Zheng, Yue and Chen, Yuhao and Qian, Bin and Shi, Xiufang and Shu, Yuanchao and Chen, Jiming},
title = {A Review on Edge Large Language Models: Design, Execution, and Applications},
year = {2025},
publisher = {Association for Computing Machinery},
url = {https://doi.org/10.1145/3719664},
journal = {ACM Comput. Surv.},
}

@article{husom2025edgeai,
author = {Husom, Erik Johannes and Goknil, Arda and Astekin, Merve and Shar, Lwin Khin and K{\aa}sen, Andre and Sen, Sagar and Mithassel, Benedikt Andreas and Soylu, Ahmet},
title = {Sustainable LLM Inference for Edge AI: Evaluating Quantized LLMs for Energy Efficiency, Output Accuracy, and Inference Latency},
year = {2025},
publisher = {Association for Computing Machinery},
url = {https://doi.org/10.1145/3767742},
journal = {ACM Trans. Internet Things},
}

@inproceedings{li2024mobilellm,
author = {Li, Xiang and Lu, Zhenyan and Cai, Dongqi and Ma, Xiao and Xu, Mengwei},
title = {Large Language Models on Mobile Devices: Measurements, Analysis, and Insights},
year = {2024},
publisher = {Association for Computing Machinery},
url = {https://doi.org/10.1145/3662006.3662059},
booktitle = {Proceedings of the Workshop on Edge and Mobile Foundation Models},
}

@misc{ouyang2025kernelbenchllmswriteefficient,
      title={KernelBench: Can LLMs Write Efficient GPU Kernels?}, 
      author={Anne Ouyang and Simon Guo and Simran Arora and Alex L. Zhang and William Hu and Christopher Ré and Azalia Mirhoseini},
      year={2025},
      eprint={2502.10517},
      archivePrefix={arXiv},
      primaryClass={cs.LG},
      url={https://arxiv.org/abs/2502.10517}, 
}

@misc{llama_cpp,
  author       = {{ggml-org contributors}},
  title        = {llama.cpp},
  year         = {2026},
  howpublished = {\url{https://github.com/ggml-org/llama.cpp}},
  note         = {GitHub repository}
}

@misc{webgpu-subgroup-matrix,
  author       = {{WebGPU Contributors}},
  title        = {“Cooperative matrix — Issue \#4195”},
  howpublished = {\url{https://github.com/gpuweb/gpuweb/issues/4195}},
  year         = {2023},
}

@software{WebKit,
  author       = {{WebKit Contributors}},
  title        = {{WebKit: WebKit Browser Engine on GitHub}},
  year         = {2025},
  url          = {https://github.com/WebKit/WebKit},
}

@misc{ruan2024webllm,
      title={{WebLLM}: A High-Performance In-Browser {LLM} Inference Engine}, 
      author={Charlie F. Ruan and Yucheng Qin and Xun Zhou and Ruihang Lai and Hongyi Jin and Yixin Dong and Bohan Hou and Meng-Shiun Yu and Yiyan Zhai and Sudeep Agarwal and Hangrui Cao and Siyuan Feng and Tianqi Chen},
      year={2024},
      eprint={2412.15803},
      archivePrefix={arXiv},
      primaryClass={cs.LG},
      url={https://arxiv.org/abs/2412.15803}, 
}

@inproceedings{lee2019ondevice,
    title	= {On-Device Neural Net Inference with Mobile {GPUs}},
    author	= {Juhyun Lee and Nikolay Chirkov and Ekaterina Ignasheva and Yury Pisarchyk and Mogan Shieh and Fabio Riccardi and Raman Sarokin and Andrei Kulik and Matthias Grundmann},
    year	= {2019},
    url = {https://doi.org/10.48550/arXiv.1907.01989},
    booktitle	= {Efficient Deep Learning for Computer Vision CVPR 2019 (ECV2019)}}

@inproceedings{nugteren2018clblast,
author = {Nugteren, Cedric},
title = {{CLBlast}: A Tuned {OpenCL BLAS} Library},
year = {2018},
isbn = {9781450364393},
url = {https://doi.org/10.1145/3204919.3204924},
booktitle = {Proceedings of the International Workshop on OpenCL},
series = {IWOCL '18}
}

@article{kerneltuner,
  author  = {Ben van Werkhoven},
  title   = {Kernel Tuner: A search-optimizing {GPU} code auto-tuner},
  journal = {Future Generation Computer Systems},
  year = {2019},
  url = {https://doi.org/10.1016/j.future.2018.08.004}
}

@inproceedings{chen2018learning,
author = {Chen, Tianqi and Zheng, Lianmin and Yan, Eddie and Jiang, Ziheng and Moreau, Thierry and Ceze, Luis and Guestrin, Carlos and Krishnamurthy, Arvind},
title = {Learning to optimize tensor programs},
year = {2018},
booktitle = {Proceedings of the 32nd International Conference on Neural Information Processing Systems},
series = {NIPS'18},
url = {https://doi.org/10.48550/arXiv.1805.08166}
}

@inproceedings{ansel2014opentuner,
author = {Ansel, Jason and Kamil, Shoaib and Veeramachaneni, Kalyan and Ragan-Kelley, Jonathan and Bosboom, Jeffrey and O'Reilly, Una-May and Amarasinghe, Saman},
title = {{OpenTuner}: an extensible framework for program autotuning},
year = {2014},
url = {https://doi.org/10.1145/2628071.2628092},
booktitle = {Proceedings of the 23rd International Conference on Parallel Architectures and Compilation},
series = {PACT '14}
}

@inproceedings{ragankelley2013halide,
author = {Ragan-Kelley, Jonathan and Barnes, Connelly and Adams, Andrew and Paris, Sylvain and Durand, Fr\'{e}do and Amarasinghe, Saman},
title = {Halide: a language and compiler for optimizing parallelism, locality, and recomputation in image processing pipelines},
year = {2013},
url = {https://doi.org/10.1145/2491956.2462176},
booktitle = {Proceedings of the 34th ACM SIGPLAN Conference on Programming Language Design and Implementation},
series = {PLDI '13}
}

@misc{pennycook2016metricperformanceportability,
      title={A Metric for Performance Portability}, 
      author={S. J. Pennycook and J. D. Sewall and V. W. Lee},
      year={2016},
      eprint={1611.07409},
      archivePrefix={arXiv},
      primaryClass={cs.PF},
      url={https://arxiv.org/abs/1611.07409}, 
}

@article{lin2025awq,
author = {Lin, Ji and Tang, Jiaming and Tang, Haotian and Yang, Shang and Xiao, Guangxuan and Han, Song},
title = {AWQ: Activation-aware Weight Quantization for On-Device LLM Compression and Acceleration},
year = {2025},
publisher = {Association for Computing Machinery},
address = {New York, NY, USA},
url = {https://doi.org/10.1145/3714983.3714987},
journal = {GetMobile: Mobile Comp. and Comm.},
}

@inproceedings{sheng2023flexgen,
author = {Sheng, Ying and Zheng, Lianmin and Yuan, Binhang and Li, Zhuohan and Ryabinin, Max and Chen, Beidi and Liang, Percy and R\'{e}, Christopher and Stoica, Ion and Zhang, Ce},
title = {FlexGen: high-throughput generative inference of large language models with a single GPU},
year = {2023},
publisher = {JMLR.org},
booktitle = {Proceedings of the 40th International Conference on Machine Learning},
url = {https://dl.acm.org/doi/abs/10.5555/3618408.3619696},
location = {Honolulu, Hawaii, USA},
series = {ICML'23}
}

@inproceedings{osama2023streamk,
author = {Osama, Muhammad and Merrill, Duane and Cecka, Cris and Garland, Michael and Owens, John D.},
title = {Stream-K: Work-Centric Parallel Decomposition for Dense Matrix-Matrix Multiplication on the GPU},
year = {2023},
publisher = {Association for Computing Machinery},
url = {https://doi.org/10.1145/3572848.3577479},
booktitle = {Proceedings of the 28th ACM SIGPLAN Annual Symposium on Principles and Practice of Parallel Programming},
location = {Montreal, QC, Canada},
series = {PPoPP '23}
}

@misc{wen2025multikernelbench,
      title={MultiKernelBench: A Multi-Platform Benchmark for Kernel Generation}, 
      author={Zhongzhen Wen and Yinghui Zhang and Zhong Li and Zhongxin Liu and Linna Xie and Tian Zhang},
      year={2025},
      eprint={2507.17773},
      archivePrefix={arXiv},
      primaryClass={cs.DC},
      url={https://arxiv.org/abs/2507.17773}, 
}

@misc{li2025tritonbenchbenchmarkinglargelanguage,
      title={TritonBench: Benchmarking Large Language Model Capabilities for Generating Triton Operators}, 
      author={Jianling Li and Shangzhan Li and Zhenye Gao and Qi Shi and Yuxuan Li and Zefan Wang and Jiacheng Huang and Haojie Wang and Jianrong Wang and Xu Han and Zhiyuan Liu and Maosong Sun},
      year={2025},
      eprint={2502.14752},
      archivePrefix={arXiv},
      primaryClass={cs.CL},
      url={https://arxiv.org/abs/2502.14752}, 
}

@misc{nvidia_cublas_2025,
  author       = {{NVIDIA Corporation}},
  title        = {{cuBLAS Library User Guide}},
  year         = {2025},
  url          = {https://docs.nvidia.com/cuda/pdf/CUBLAS\_Library.pdf},
  note         = {Accessed: 2026-04-08}
}

@inproceedings{hendrycks2021mmlu,
  author = {Hendrycks, Dan and Burns, Collin and Basart, Steven and Zou, Andy and Mazeika, Mantas and Song, Dawn and Steinhardt, Jacob},
  title = {Measuring Massive Multitask Language Understanding},
  booktitle = {International Conference on Learning Representations (ICLR)},
  year = {2021},
  url = {https://doi.org/10.48550/arXiv.2009.03300}
}

@misc{grattafiori2024llama3,
  title         = {The {Llama} 3 Herd of Models},
  author        = {Aaron Grattafiori and Abhimanyu Dubey and Abhinav Jauhri and others},
  year          = {2024},
  eprint        = {2407.21783},
  archiveprefix = {arXiv},
  primaryclass  = {cs.AI},
  url           = {https://arxiv.org/abs/2407.21783}
}

@misc{yang2025qwen3,
  author = {Yang, An and others},
  title = {Qwen3 Technical Report},
  year = {2025},
  eprint = {2505.09388},
  archivePrefix = {arXiv},
  primaryClass = {cs.CL},
  url = {https://doi.org/10.48550/arXiv.2505.09388}
}

@inproceedings{chen2025weinfer,
  title = {{WeInfer}: Unleashing the Power of {WebGPU} on {LLM} Inference in Web Browsers},
  author = {Zhiyang Chen and Yun Ma and Haiyang Shen and Mugeng Liu},
  booktitle = {Proceedings of the ACM Web Conference 2025 (WWW)},
  year = {2025},
  url = {https://doi.org/10.1145/3696410.3714553}
}

@inproceedings{jia2024nnjit,
  title = {Empowering In-Browser Deep Learning Inference on Edge Through Just-In-Time Kernel Optimization},
  author = {Jia, Xiang and Jiang, Pinming and Wang, Jiayu and Fu, Fangcheng and Miao, Xupeng and Nie, Xiaonan and Cui, Bin},
  booktitle = {Proceedings of the 22nd Annual International Conference on Mobile Systems, Applications, and Services (MobiSys)},
  year = {2024},
  url = {https://doi.org/10.1145/3643832.3661892}
}

@inproceedings{cho2023gptunecrowd,
  author    = {Younghyun Cho and James Weldon Demmel and Jacob King and Xiaoye S. Li and Yang Liu and Hengrui Luo},
  title     = {Harnessing the Crowd for Autotuning High-Performance Computing Applications},
  booktitle = {IEEE International Parallel and Distributed Processing Symposium (IPDPS)},
  pages     = {635--645},
  year      = {2023},
  url = {https://ieeexplore.ieee.org/document/10177466},
  doi       = {10.1109/IPDPS54959.2023.00069}
}

@inproceedings{xu2024npu,
author = {Xu, Daliang and Zhang, Hao and Yang, Liming and Liu, Ruiqi and Huang, Gang and Xu, Mengwei and Liu, Xuanzhe},
title = {Fast On-device {LLM} Inference with {NPUs}},
year = {2025},
booktitle = {Proceedings of the 30th ACM International Conference on Architectural Support for Programming Languages and Operating Systems (ASPLOS)},
doi = {10.1145/3669940.3707239},
url = {https://doi.org/10.1145/3669940.3707239}
}

@misc{xue2024powerinfer2,
title = {{PowerInfer-2}: Fast Large Language Model Inference on a Smartphone},
author = {Zhenliang Xue and Yixin Song and Zeyu Mi and Xinrui Zheng and Yubin Xia and Haibo Chen},
year = {2024},
eprint = {2406.06282},
archivePrefix = {arXiv},
primaryClass = {cs.DC},
url = {https://arxiv.org/abs/2406.06282}
}

@INPROCEEDINGS{karami2025perfhorizons,
  author={Karami, Rachid and Kao, Sheng-Chun and Kwon, Hyoukjun},
  booktitle={2025 IEEE International Symposium on Performance Analysis of Systems and Software (ISPASS)},
  title={Understanding the Performance Horizon of the Latest ML Workloads with NonGEMM Workloads},
  year={2025},
  url={https://doi.org/10.48550/arXiv.2404.11788}
}

@INPROCEEDINGS{nugteren2015cltuner,
  author={Nugteren, Cedric and Codreanu, Valeriu},
  booktitle={2015 IEEE 9th International Symposium on Embedded Multicore/Many-core Systems-on-Chip},
  title={{CLTune}: A Generic Auto-Tuner for {OpenCL} Kernels},
  year={2015},
  url={https://doi.org/10.1109/MCSoC.2015.10}
}

@INPROCEEDINGS{sorensen2019onesize,
  author={Sorensen, Tyler and Pai, Sreepathi and Donaldson, Alastair F.},
  booktitle={2019 IEEE International Symposium on Workload Characterization (IISWC)},
  title={One Size Doesn't Fit All: Quantifying Performance Portability of Graph Applications on {GPUs}},
  year={2019},
  pages={155-166},
  url={https://doi.org/10.1109/IISWC47752.2019.9042139}
}

@InProceedings{zhang2013perfport,
author={Zhang, Yao and Sinclair, Mark and Chien, Andrew A.},
title={Improving Performance Portability in {OpenCL} Programs},
booktitle={Supercomputing},
year={2013},
url={https://doi.org/10.1007/978-3-642-38750-0\_11}
}

@inproceedings{cudadma,
author = {Bauer, Michael and Cook, Henry and Khailany, Brucek},
title = {CudaDMA: optimizing GPU memory bandwidth via warp specialization},
year = {2011},
isbn = {9781450307710},
publisher = {Association for Computing Machinery},
address = {New York, NY, USA},
url = {https://doi.org/10.1145/2063384.2063400},
doi = {10.1145/2063384.2063400},
booktitle = {Proceedings of 2011 International Conference for High Performance Computing, Networking, Storage and Analysis},
articleno = {12},
numpages = {11},
location = {Seattle, Washington},
series = {SC '11}
}

@manual{cuda_guide_12_9,
  title        = "{CUDA C++ Programming Guide}",
  author       = "{NVIDIA Corporation}",
  organization = "{NVIDIA Corporation}",
  edition      = "Release 12.9",
  address      = "Santa Clara, CA",
  year         = {2025},
  month        = jun,
  url          = {https://docs.nvidia.com/cuda/pdf/CUDA\_C\_Programming\_Guide.pdf},
  note         = {Section~7.1, p.~140}
}

@inproceedings{liu2021gptune,
author = {Liu, Yang and Sid-Lakhdar, Wissam M. and Marques, Osni and Zhu, Xinran and Meng, Chang and Demmel, James W. and Li, Xiaoye S.},
title = {GPTune: multitask learning for autotuning exascale applications},
year = {2021},
isbn = {9781450382946},
publisher = {Association for Computing Machinery},
address = {New York, NY, USA},
url = {https://doi.org/10.1145/3437801.3441621},
doi = {10.1145/3437801.3441621},
booktitle = {Proceedings of the 26th ACM SIGPLAN Symposium on Principles and Practice of Parallel Programming},
pages = {234–246},
numpages = {13},
location = {Virtual Event, Republic of Korea},
series = {PPoPP '21}
}

@inproceedings{song2024powerinfer,
author = {Song, Yixin and Mi, Zeyu and Xie, Haotong and Chen, Haibo},
title = {PowerInfer: Fast Large Language Model Serving with a Consumer-grade GPU},
year = {2024},
isbn = {9798400712517},
publisher = {Association for Computing Machinery},
address = {New York, NY, USA},
url = {https://doi.org/10.1145/3694715.3695964},
doi = {10.1145/3694715.3695964},
booktitle = {Proceedings of the ACM SIGOPS 30th Symposium on Operating Systems Principles},
pages = {590–606},
numpages = {17},
location = {Austin, TX, USA},
series = {SOSP '24}
}

@inproceedings{alizadeh2024llminflash,
author = {Alizadeh, Keivan and Mirzadeh, Iman and Belenko, Dmitry and Khatamifard, Karen and Cho, Minsik and Del Mundo, Carlo C. and Rastegari, Mohammad and Farajtabar, Mehrdad},
title = {{LLM} in a Flash: Efficient Large Language Model Inference with Limited Memory},
year = {2024},
booktitle = {Proceedings of the 62nd Annual Meeting of the Association for Computational Linguistics (ACL)},
eprint = {2312.11514},
archivePrefix = {arXiv},
url = {https://arxiv.org/abs/2312.11514}
}

@article{wang2024anatomizing,
author = {Wang, Qipeng and Jiang, Shiqi and Chen, Zhenpeng and Cao, Xu and Li, Yuanchun and Li, Aoyu and Ma, Yun and Cao, Ting and Liu, Xuanzhe},
title = {Anatomizing Deep Learning Inference in Web Browsers},
journal = {ACM Transactions on Software Engineering and Methodology},
year = {2025},
issue_date = {February 2025},
publisher = {Association for Computing Machinery},
address = {New York, NY, USA},
volume = {34},
number = {2},
issn = {1049-331X},
url = {https://doi.org/10.1145/3688843},
doi = {10.1145/3688843},
month = jan,
articleno = {47},
numpages = {43}
}

@article{nadaraya1964estimating,
  author    = {Nadaraya, E. A.},
  title     = {On Estimating Regression},
  journal   = {Theory of Probability and Its Applications},
  volume    = {9},
  number    = {1},
  pages     = {141--142},
  year      = {1964},
  doi       = {10.1137/1109020},
  url       = {https://doi.org/10.1137/1109020}
}

@misc{levine2026llamaweb,
  author       = {Reese Levine and Rithik Sharma and Nikhil Jain and Abhijit Ramesh and Zheyuan Chen and Neha Abbas and James Contini and Tyler Sorensen},
  title        = {Llamas on the Web: Memory-Efficient, Performance-Portable, and Multi-Precision {LLM} Inference with {WebGPU}},
  howpublished = {arXiv preprint arXiv:2605.20706},
  year         = {2026},
  url          = {https://arxiv.org/abs/2605.20706}
}

\end{document}